# On the nature of intraspecific genetic variability:
# Evidence against the ruling paradigm

Anastassia M. Makarieva*, Victor G. Gorshkov

Theoretical Physics Division, Petersburg Nuclear Physics Institute, 188300, Gatchina, St. Petersburg, Russia, elba@peterlink.ru, http://www.bioticregulation.ru

*Corresponding author

*Summary*

Empirical evidence is presented which contradicts the established interpretation of the intraspecific genetic variability as the adaptive potential of the species: the uniform evolutionary tempo across the life kingdom, species discreteness, and absence of correlation between genetic variability and prosperity of extant species testify against the ruling paradigm. Consistent interpretation of the nature of intraspecific genetic variability is based on recognizing the limited sensitivity of stabilising selection, which allows for accumulation and persistence in the population of a considerable amount of mutational substitutions, which, to some degree, erase the meaningful genetic information of the species. The proposed interpretation also provides solution to the inbreeding paradox in the invasive species.

*1. Introduction: Variability as adaptive potential, major lines of logic and evidence*

The statement that intraspecific genetic variability represents the species' adaptive potential represents one of the ruling biological paradigms. It is based on the following consideration: the more genetic variants are present in the population, the more chances the population has of surviving if the environment changes.

The connection of this qualitative idea to empirical evidence seems to be most obvious in the realm of artificial selection. On the basis of intraspecific genetic variability man selects for animals and plants with particular useful properties (e.g., high milk production). If we consider this selective pressure as a sort of "changing environment", this means that only those populations survive which feature high variability comprising the properties that are selected for. Individuals deprived of the needed properties are artificially eliminated by man ("external conditions").

There is further a handful of other facts conventionally interpreted in favour of the paradigm, like the textbook examples of industrial melanism or sickle cell anaemia. In the first case, in the population of butterflies that are polymorphic with respect to colour (dark- and light-coloured individuals), dark individuals are considered to be better adapted to the environment of the industrial regions where the ground surface is dirty and dark, as they are less conspicuous at this background than their light-coloured conspecifics. Sickle cell anaemia, a serious genetic disorder in humans when homozygous, imparts to its heterozygous carriers some resistance against malaria, another dangerous disease. The presumably adaptive nature of this polymorphism is involved to explain the elevated frequency of sickle cell anaemia in those regions of Africa that are most affected by malaria.

Finally, the significance of the intraspecific genetic variability for speciation is illustrated on the examples of the so-called ring species, like, e.g., the gulls *Larus fuscus* and *Larus argentatus*. These birds form a set of populations living around the Arctic ocean (Mayr, 1963; Green *et al*., 1989). The lesser black-backed gull *L. fuscus* is a common species in Western Europe. Its range extends east into the Russian Arctic through populations that are interbreeding but which can be arranged into several subspecies, each slightly different. The easternmost subspecies is so far east that it ultimately ranges again into Western Europe as the herring gull, where it exists alongside with the lesser black-backed gull. But the herring gull does not interbreed with the lesser black-backed gull in Western Europe and is called there *Larus argentatus*. These observations are interpreted in that sense that the circumpolar environmental gradient, when acting upon the intraspecific genetic variability of *L. fuscus*, produced a new species, *L. argentatus*.

Such arguments, which, for over a century now, have been learnt by heart by every biology student, are considered to be sufficient to prove the adaptive significance of the intraspecific genetic variability beyond any further doubt. As discussed below, the evidence contradicting this paradigm is either ignored or considered as a paradox, to which any possible explanation can be sought for, but not the one questioning the paradigm itself.

*2. Evidence contradicting the paradigm*

Genetic variability is traditionally measured in relative units — heterozygosity $H$ (the mean proportion of genetic loci by which two haploid genotypes randomly picked up from the population differ from each other) or polymorphism $P$ (the proportion, among all loci studied, of variable loci found in the population). Both $H$ and $P$ are always confined between zero and unity.

However, natural selection acts upon, and selects among, individuals, not genetic loci. Following the same logic that supports the adaptive potential of variability, it is clear that it is the number of variable individuals that should matter for adaptation, not the number of variable loci per se. For example, if there is a population consisting of $N = 2$ individuals, whatever is the number of loci at which they differ, the number of different genotypes to be acted upon by natural selection is only two. Generally, given that there are four different nucleotides in the DNA, the number of possible genotypes for a population with nucleotide heterozygosity $H$ and genome size $G$ is given by $4^{GH}$. For the smallest eukaryote genome $G \sim 10^7$ base pairs and a conservative nucleotide heterozygosity $H \sim 10^{-4}$ (Li & Sadler, 1991), the number of different genotypes is already $10^{200}$, greatly exceeding the number of individuals in any population. This means that there are no genetically identical individuals within any species. Hence, the number of different genotypes present in the population is simply equal to the population size and, within broad limits, is independent of heterozygosity. The maximum number of alleles of each particular gene locus that can be simultaneously present in the population is also equal to population size.

Thus, if the intraspecific genetic variability represented the evolutionary adaptive potential of biological species, then, as far as in absolute terms it is highest in the largest populations, species featuring the greatest number of individuals should be able to adapt and evolve more rapidly than species consisting of a small number of individuals — the probability of finding an individual fitted to a particular environment is apparently higher when the choice is made among many, rather than few, genetically different individuals.

The available paleoevidence, however, strongly contradicts this prediction, Table 1. For example, in spite of the ten- to 10,000-fold difference in global species population numbers between rodents and mammalian carnivores, new species within both taxa appear every one-two million years. At least a billion-fold difference exists between the global species population numbers of lizards and microscopic marine organisms (diatoms, dinoflagellates and foraminifers), yet organisms from these taxonomic groups all speciate every twenty million years on average. Neither there is any considerable difference between speciation rates of insects and mammals, despite at least a million-fold difference in abundances, Table 1.

Although it is paradigmatically stated that more variable populations should be adapting and adapted better than less variable ones, no quantitative tests are proposed to check for the expected differences in the adaptive potential. Indeed, in what measurable variables should the expected better adaptation, associated with higher variability, be manifested? As we have shown on the basis of the speciation tempo data, Table 1, intraspecific genetic variability does not have any measurable effect on the ability of the various taxonomic groups to produce new species in the course of evolution.

The most important observation that has been made with regard to the organization of the living world, namely that biological species, both extinct and extant, are *discrete*, also testifies against the adaptive potential of intraspecific variability. Since different genotypes of one and the same species represent a *continuum* of morphological forms, evolution based on the observed intraspecific variability would have been gradual, i.e. the ancient species would have been gradually intermingling into their evolutionary successors, instead of appearing in the chronological record in the form of discrete morphological and genetic entities and persisting without any directional change during most part of the species' life span as most species do (Stanley, 1979; Jackson, 1990; Gould & Eldridge, 1993).

There are neither any indications that elevated genetic variability contributes to prosperity of the extant species or that reduced genetic variability prevents from being prosperous. For example, within the class of insects the haplodiploid order Hymenoptera (bees, ants, wasps) is, together with the diploid Coleoptera, Lepidoptera and Diptera, one of the four most species-rich and widely-spread orders with over 100,000 described species. At the same time, allozyme heterozygosity in the haplodiploid Hymenoptera is almost three times lower than in the diploid orders. It is equal to $H = 0.05 \pm 0.001$ ($\pm$ 1 s.e., $n = 64$ species), compared to $H = 0.13 \pm 0.02$ ($n = 15$) in beetles and $H = 0.14 \pm 0.001$ ($n = 62$) in butterflies (data from Nevo *et al*. (1984) and Graur (1985) analysed by Gorshkov *et al*. (2000)).

In mammals, too, there is a wealth of examples of perfectly prosperous species with negligibly low variability (Merola, 1994), as well as many cases where endangered species feature very high variability. For example, among the 321 mammalian species with known allozyme heterozygosity studied by Makarieva (2001), Fig. 1, 42 species are characterised by allozyme heterozygosity not exceeding $H = 0.01$, which is five times lower than the mammalian average, $H = 0.05$. Among these, there are many widely spread species like rodents *Dipodomys*, *Peromyscus*, *Spermophilus*, *Rattus*, which can be in no way characterised as endangered. Among the non-rodent species, some animals with low heterozygosity are indeed endangered like the famous cheetah *Acinonyx jubatus*, but others are quite numerous like, for example, the northern elephant seal *Mirounga angustirostris* which, at the time of heterozygosity measurements, numbered over 30,000 individuals. This led the researchers to conclude that "genic variability is not essential for the continued existence of animal species" (Bonnel & Selander, 1974). Similar remarks with respect to other taxa are not uncommon in the literature, see, e.g., Bates & Zink (1982), but largely remain ignored as they contradict the ruling paradigm. In the meantime, the critically endangered greater one-horned rhinoceros *Rhinoceros unicornis* exhibits a high allozyme heterozygosity, $H = 0.10$, twice the mammalian mean (Dinerstein & McCracken, 1990). Populations of the Przewalski horse feature the highest allozyme heterozygosity ever recorded in mammals ($H \sim 0.4$) and are characterised by high juvenile mortality and decreased lifespan (Bowling & Ryder, 1987).

With the on-going elimination of natural ecosystems in the course of civilisation growth, the majority of the natural species affected either go extinct or are driven on the verge of extinction. A few species, however, proved to be able to thrive in the anthropogenically transformed environments — these are, for example, urban cats, pigeons, sparrows, domestic rats, mice, cockroaches etc. Is there any evidence that these animals, viewed as successfully adapted within the traditional paradigm, possess some extraordinarily high intraspecific variability that should have presumably allowed them to flourish in the changed environment, in contrast to other natural species who did not survive the human interference into their lifestyle? According to the available evidence, the answer is a unambiguous no, Table 2. All such species appear to be characterized by heterozygosity values well within the taxonomic range. The German cockroach *Blatella germanica*, which, following the humans, invaded the entire planet, posesses a vanishingly small variability, see Table 2 and Cloarec *et al*. (1999).

It can be concluded that the dogma of the indispensability of high intraspecific variability for adaptation and evolution does not reside on whatsoever quantitatively consistent empirical evidence but, rather, is supported by the long-standing tradition *per se* and has been sustained by the apparent lack of scientific efforts to find a non-contradictory explanation for the observed patterns.

*3. Non-contradictory explanation of the intraspecific genetic variability*

Organisms live and function in their natural environment, with which their morphological and behavioural properties are rigidly correlated. Genetic information about the appropriate interaction with the environment is written into the species' genome. When a new individual is produced, the parental genetic information has to be copied. During copying, some misprints (mutations) inevitably accumulate that erase the meaningful information of the normal genome. It is well-known that the worst of these misprints are eliminated by the stabilising selection. However, it is traditionally held that, by definition, mutations sustained in the population are not very harmful for individuals and, hence, can represent their adaptive potential. The non-contradictory explanation, instead, consists in admitting the limited sensitivity of the stabilising selection.

Every published book contains a certain number of misprints which nevertheless do not prevent the reader from wholly grasping the book's content. Similarly, a certain number of mutations can accumulate in the genome without being "seen" by the stabilising selection. Individuals with genotypes differing from the normal species genome by less than a critical number $n_C$ of mutational substitutions, which characterizes the sensitivity of selection, are equally competitive in the population. However, similar to misprints, these mutations are all harmful for the species, representing the permissible level of erosion of the species' genetic information.

Under distorted environmental conditions, most properties that in the natural environment impart high competitive capacity to their carriers, appear useless. No longer eliminated by the stabilising selection, mutations start to accumulate in the genome beyond $n_C$. The intraspecific genetic variability grows. (Similarly, books published without a single misprint are all equivalent, while the more misprints, the more different from each other the individual books become, but acquire no additional information.) The accumulation of mutations beyond $n_C$ can occur up to the lethal threshold $n_L$. Individuals with $n > n_L$ mutational substitutions in their genome are either lethal or infertile. When the environmental conditions degrade even further, the normal genome completely loses its meaning; the species undergoes gradual genetic degradation; the frequency of various genetic malformations increases.

Genetic degradation of species in the distorted environment is a random chaotic process, when different genotypes with $n > n_C$ succeed each other approaching the lethal threshold. When selecting among the various malformations featured by individuals with $n > n_C$, man can occasionally find some properties that, while prohibitive for the existence in the natural environment, can appear useful for man, e.g., the very large udders in the domestic cow, which impede normal locomotion of the animal and would make it vulnerable to attacks of predators in the wild. Thus, artificial selection does not lead to appearance of new meaningful genetic information, but, rather, represents a bizarre (from the nature's point of view) choice of particular malformations featured by individuals that have lost information about how to interact correctly with the environment and are genetically balancing on the verge of inviability. This explains why, in spite of the fact that the genetic variability of artificially selected domestic mammals, Fig. 1, greatly exceeds the minimum genetic distance between close mammalian species (Avise & Aquadro, 1982), no new species has ever originated in the course of artificial selection.

While the species is genetically degrading and ultimately goes extinct in the distorted environment, in the succession of genotypes with $n > n_C$ some are eliminated later than the others. When a population of butterflies is degrading in the urban environment under the additional pressure of carnivorous urban birds, the less-conspicuous, dark-coloured individuals can survive longer than the light-coloured ones. However, this does mean that the dark-coloured individuals have adapted better to the urban environment — they are simply the last to perish. This "adaptation" has nothing to do with the ability of natural species to sustainably exist in their natural environments for millions of years.

In very much the same manner as random misprints cannot transform the first book volume into the second one, mutations erasing the genetic information of the species cannot facilitate biological evolution, i.e. change of the meaningful genetic information. It is not surprising therefore that, as discussed in Section 2, the evolutionary tempo is not affected by the amount of intraspecific genetic variability. At the same time, within the proposed framework it is easy to explain the ring species phenomenon.

Consider two populations A and B of a species with the sensitivity of competitive interaction equal to $n_C$, when individuals tolerate $n \leq n_C$ mutational substitutions in their genomes without loss of competitive capacity. While the total number of mutational substitutions in all populations of the species is, on average, the same, their localisation is, in the general case, different. The number of mutational substitutions $n_{AB}$ in the offspring produced by two individuals, one from population A and another from population B, is therefore on average confined between $n_C$, when all mutational substitutions in the two populations are identical, and $2n_C$, when they are all different, Fig. 2, $n_C \leq n_{AB} \leq 2n_C$. In the latter case, when $n_{AB}$ reaches its maximum, it can go beyond the lethal threshold $n_L$, $n_{AB} \geq n_L$, and the progeny of individuals from two distant isolated populations will be inviable or infertile, making successful interbreeding impossible. This fact, however, contrary to the traditional interpretation, does not indicate formation of two new species A and B. Generally, the following criterium can be formulated: if one genotype can be gradually turned to another genotype via a succession of viable forms, both genotypes belong to the same species. The notion of subspecies, to which no formal definition exists, most accurately corresponds to distant populations, hybridization between which yields inviable or infertile offspring, like the populations of *Larus fuscus* and *L. argentatus*.

*4. Inbreeding and genetic variability*

In diploid organisms, mutations that affected a particular locus in the one copy of the genome can be partially or completely masked by the proper functioning of the intact, mutation-free locus of the second copy. Thus, the number $n_C$ of masked mutational substitutions that can be tolerated in a diploid population should be much larger than the number of mutations tolerated in a haploid population, where all mutations are manifested. For example, allozyme heterozygosity of haplodiploid insects (hymenopterans and thrips) is much lower than that of other insect orders (Graur, 1985; Crespi, 1991). Similarly, human autosomes that are always diploid are much more variable than sex chromosomes that are effectively haploid (hemizygous) in males. Whereas approximately 1 in 560 bp is variant in the autosomal human DNA, the variability of the X-chromosomal DNA is only about 1 in 2,100 bp, while the variability of the Y chromosome (which is never diploid) is less than 1 in 48,000 bp (Hofker *et al.*, 1986; Jakubiczka *et al.*, 1989; Malaspina *et al.*, 1990; Dorit *et al.*, 1995).

When closely related individuals interbreed, there is a high probability that their offspring become homozygous with respect to many mutational substitutions that remained heterozygous and, hence, masked in the parents. In the result, while the heterozygosity of the offspring, obviously, decreases, the number $n$ of unmasked deleterious mutational substitutions in the offspring grows dramatically beyond $n_C$ and towards the lethal threshold $n_L$. Therefore, among the inbred offspring there should be many individuals with significantly reduced biological performance (incompetitive, susceptible to diseases, having reduced life span etc.). It is clear that inbreeding should produce most negative effects in populations that have for a long time existed under unnatural conditions and, having accumulated a high number of deleterious substitutions, feature high genetic variability, like domestic or laboratory populations (e.g., *Drosophila*). In *Homo sapiens*, the observed negative consequences of inbreeding brought about the various cultural taboes on marriages of close relatives. The necessity of these taboes testifies for the fact that in humans, like in all other animals, there is no genetically encoded predisposition against inbreeding.

Thus, the inbreeding-related reduced biological performance has nothing to do with the loss of genetic variability *per se*, contrary to the explanation of this effect within the traditional paradigm. As discussed in Section 2, many species are perfectly prosperous with negligible variability. (It should also be stressed that the low genetic variability in natural populations is not at all necessarily caused by inbreeding or bottleneck effects, but owes itself to the very sensitive process of natural selection, which, in some species, does not permit even the slightest degree of erosion of the normal genome.) Instead, the various malformations appearing in the course of inbreeding are due to the fact that the deleterious variation masked in the parental genomes becomes unmasked when homozygous in the inbred offspring. These effects once again prove the inherently deleterious nature of the intraspecific genetic variability. Remarkably, consistent with our explanation but unexplained within the traditional paradigm, there are no cases of inbreeding depression reported for haploid species.

The inbreeding coin has also an opposite, positive side completely ignored in the traditional consideration of this phenomenon. When the inbred offspring become homozygous, some of them become homozygous with respect to the deleterious mutational substitutions, while the others become homozygous with respect to the normal genome. Thus, inbreeding serves to purify the species genome, as it produces, along with individuals with various malformations ("monsters"), also individuals completely freed from whatever genetic defects ("geniuses").

Another way of unmasking and eliminating the deleterious substitutions of the diploid genome is to turn it into the haploid form. Such mechanisms of genome purification are widely used in nature. Due to the dramatic increase of $n$ beyond $n_C$ and high rates of elimination of incompetitive haploid individuals, the population numbers in the haplophase should be much larger than in the diplophase, to ensure that normal haploid individuals with $n < n_C$ are always present. Accordingly, the number of spermatozoids in the haplophase of vertebrate species exceeds the number of diploid individuals by hundred million times. Similarly, the number of haploid males in the haplodiploid insect species is much larger than the actual number of males needed to fertilise the queen.

The proposed interpretation of the inbreeding phenomenon resolves the so-called inbreeding paradox in the invasive species (Allendorf & Lundquist, 2003; Frankham, 2005; Pérez et al., 2005a,b). Indeed, within the traditional paradigm it is impossible to explain how the invasive species, which by definition appear in the alien ecosystem in small numbers and, hence, are depleted of genetic variation, manage to "adapt" to the new environment. Provided that the founding couple produces a large enough number of normal offspring (homozygous with respect to the normal genome), the following generations of the invasive species are no less competitive than their conspecifics from the original population in the natural environment. The invasion success, apparently unrelated to genetic variability, depends instead on the generality of the natural habits of the invasive species. What is, from the antropocentric point of view, adaptation of an invasive species to a new environment, can be, from the point of view of the invasive species itself, just an extension of its range with no noticeable environmental change, if this species is genetically encoded to depend on but a few general environmental parameters.

*5. Conclusions*

We have argued that intraspecific genetic variability arises due to the inevitable erosion of the meaningful genetic information of the species during its copying. Genetic variability further persists in the population due to the limited sensitivity of the process of stabilising selection, which does not "notice" the deleterious changes until they accumulate in the genome in appreciable amounts. Evidence was presented that, in accordance with the proposed interpretation of genetic variability and contrary to its traditional consideration as species' adaptive potential, lack of genetic variation does not have any effect on the evolutionary success of biological species.

In the more general framework, the idea that in the course of evolution species have to adapt to the changing environment, is challenged by the rapidly accumulating evidence that the environment on Earth is under control of the biota itself, see, e.g., (Gorshkov *et al*., 2000; 2004). If there is an optimal environment that has to be sustained, the program of how to do it should be written in the normal genomes of species composing the natural ecological community. If the environment changes, the species of the ecological community should initiate actions aimed at its recovery to the optimum on the basis of their genetic program, instead of starting to change genetically themselves. Hence, all facts testifying in favour of the biotic nature of environmental stability on Earth represent, in their essence, additional arguments against the adaptive potential of intraspecific genetic variability. Biodiversity can be saved in no other way than via the restoration and conservation of the natural ecosystems.

**Table 1. Species duration versus global species population numbers in different organisms (after Makarieva & Gorshkov, 2004).**

| Taxonomic group | Species duration, Myr | Global species abundance, individuals |
|---|---|---|
| Marine diatoms | 8-25 | $10^{18}$ |
| Dinoflagellates | 16 | $10^{17}$ |
| Planktic foraminifers | > 20 | $10^{17}$ |
| Benthic foraminifers | > 20 | >$10^{15}$ |
| Insects (beetles, *Drosophila*) | > 2 | $10^{11}$–$10^{13}$ |
| Higher plants (pine trees) | > 8 | $10^{11}$–$10^{12}$ |
| Bryophytes | >20 | $10^{16}$ |
| Lizards | 26 | $10^5$–$10^9$ |
| Turtles, crocodiles | 5 | $10^5$ |
| Rodents | 1 | $10^5$–$10^8$ |
| Carnivores | 1.2 | $10^4$ |

Major data sources: speciation rate (Stanley, 1985; Bush *et al*., 1977); population size data (Nei & Graur, 1984); analysis by Makarieva & Gorshkov (2004).

**Table 2. Allozyme heterozygosity *H* in species thriving in anthropogenically-transformed environments (but not directly exposed to artificial selection)**. *L* is the number of allozyme loci studied. Mean *H* for Aves and Insecta (excluding *Drosophila*) are taken from Nevo *et al.* (1984), for Mammalia from Makarieva (2001).

| Species | *H* | *L* | Source | Taxon mean *H* (± 1 s.d.) |
|---|---|---|---|---|
| German cockroach *Blatella germanica* | 0.012 | 19 | Nevo *et al*. (1984) | 0.09 ± 0.06 (Insecta) |
| Common pigeon *Columba livia domestica* | 0.075 | 22 | Kimura & Yamamoto (1982) | 0.05 ± 0.03 (Aves) |
| House sparrow *Passer domesticus* | 0.029 | 29 | Bates & Zink (1992) | 0.05 ± 0.03 (Aves) |
| Domestic cat *Felis catus* | 0.07 | 56 | O'Brien (1980) | 0.05 ± 0.04 (Mammalia) |
| Domestic rat *Rattus norvegicus* | 0.064 | 25 | Nevo *et al*. (1984) | 0.05 ± 0.04 (Mammalia) |
| House mouse *Mus musculus* | 0.066 | 33 | Nevo *et al*. (1984) | 0.05 ± 0.04 (Mammalia) |

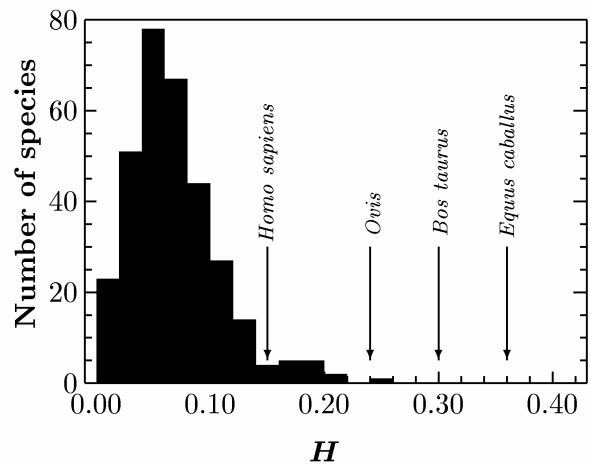

**Fig. 1. Frequency distribution of allozyme heterozygosity *H* in natural (non-zoo, non-domestic, non-laboratory) mammalian species (*n* = 321) (data of Makarieva (2001)).** Arrows indicate heterozygosity values of the domestic animals (horses *Equus caballus* (Bowling & Ryder, 1987); cow *Bos taurus* (Bannikova & Zubareva, 1995); domestic sheep *Ovis* (Wang *et al.*, 1990)) and man (Nevo *et al.*, 1984).

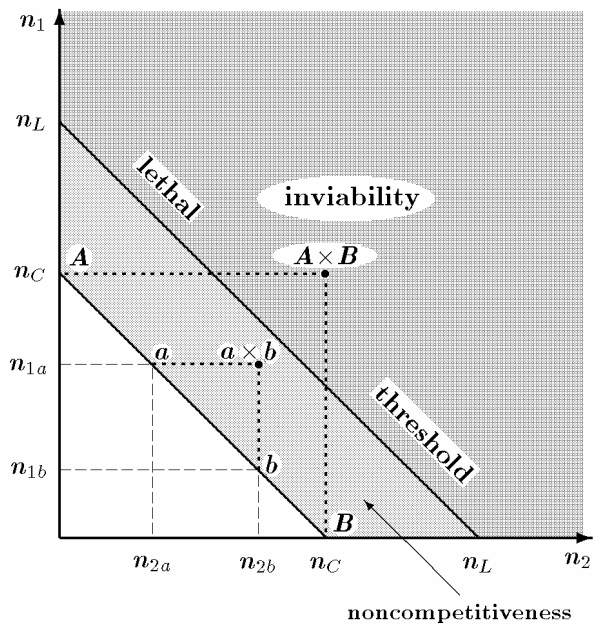

**Fig. 2. Genetic differences between different populations and subspecies of a single species (Gorshkov *et al.* (2000)).**
Let population A be the reference population. The species genome $G$ is divided in a mosaic fashion into two equal parts $G_1$ and $G_2$ in such a manner that all $n_C$ mutational substitutions (deviations from the normal genome) encountered in population A are located in the part $G_1$, whereas the part $G_2$ is free from decay substitutions. Dictated by the species-specific sensitivity of stabilising selection, the mean number of mutational substitutions in individuals in each population is close to $n_C$, but their localisation in isolated populations is different. Let $n_1$ and $n_2$ be the number of substitutions in the $G_1$ and $G_2$ parts of the genome, respectively. While in population A all $n$ substitutions are located in the part $G_1$ of the species genome, $n = n_1 = n_C$, in other populations some mutational substitutions ($n_1$) are located in the $G_1$ part, while others ($n_2$) in the $G_2$ part of the species genome, $n = n_1 + n_2 = n_C$. Finally, there may be a population B where all the $n_C$ mutational substitutions are located in the part $G_2$, $n = n_2 = n_C$. Hence, all populations and subspecies are described by a straight line AB, each point of which corresponds to some particular localisation of decay substitutions in the genome. The genetic equivalence of all populations and subspecies is manifested by the fact that they are all equally close to the normal genome, the measure of distance being $n_C$. The line parallel to AB corresponds to the lethal threshold $n_L$. Hybridisation of individuals from two isolated populations a and b (point a×b) drives the number of mutational substitutions in the offspring beyond the border AB ($n > n_C$). Such offspring are viable but noncompetitive as compared to normal individuals of both subspecies. Hybridisation of distant subspecies A and B (point A×B) yields inviable offspring. The difference between isolated populations and subspecies is purely quantitative.